\documentclass[useAMS,usenatbib]{mn2e}

\usepackage {graphicx}

\title[The NLTE Barium Abundance in Dwarf Stars]
{The NLTE Barium Abundance in Dwarf Stars in the Metallicity Range of 
-1$<$ [Fe/H] $<$+0.3}
\author[S.~Korotin et al.]{S.~Korotin$^{1}$\thanks{E-mail:serkor@skyline.od.ua},
T.~Mishenina$^{1}$, T.~Gorbaneva$^{1}$, and C.~Soubiran$^{2}$
\\
$^{1}$Astronomical Observatory of Odessa National University,
and Isaac Newton Institute of Chile Odessa branch,
Odessa, Ukraine\\
$^{2}$Universit\'e Bordeaux 1, CNRS - Laboratoire d'Astrophysique de Bordeaux, 
UMR 5804, 33370 Floirac, France.}

\begin{document}

\date{Accepted. Received ; in original form }

\pagerange{\pageref{firstpage}--\pageref{lastpage}} \pubyear{2011}

\maketitle

\label{firstpage}

\begin{abstract}

We present the results of determination of the barium abundance considering 
the non-LTE (NLTE) effects in 172 dwarf stars in the metallicity range of 
--1$<$ [Fe/H] $<$+0.3, assigned to different Galactic substructures by kinematic 
criteria. We used a model of the Ba atom with 31 levels of Ba\,{\sc i} and 101 
levels of Ba\,{\sc ii}. The atmosphere models for the investigated stars were 
computed using the ATLAS9 code modified by new opacity distribution functions 
by \citet{ck04}.
The NLTE profiles of the unblended Ba\,{\sc ii} (4554~\AA, 5853~\AA, 6496~\AA) 
were computed and then compared to those observed. The line 6141~\AA\ was also 
used, but with an allowance for its correlation with the iron line. The average
barium abundances in the thin and thick discs are 0.01 $\pm$0.08 and 
--0.03 $\pm$0.07, respectively. The comparison to the calculations of the 
Galactic chemical evolution by \cite{ser09} was conducted. The trend obtained 
for the Ba abundance versus [Fe/H] suggests a complex barium production process
in the thin and thick discs.

\end{abstract}


\section{Introduction}

Barium is one of the elements formed by the neutron capture processes; 
it is an important component in nucleosynthesis by the neutron capture 
reactions, as well as in reconstruction of its evolution in the Galaxy 
and the Galactic chemical evolution as a whole. 
The neutron capture reactions can proceed in two ways, depending on the 
neutron flux density: as a {\it slow }process (s-process) or as a {\it rapid }
process (r-process). 
Based on the classical model describing the solar abundance distribution of 
s-process elements with three exponential distributions of neutron exposures 
\citep{kap89}, the s-process is divided into three components: the {\it main } 
s-component is produced in the asymptotic giant branch (AGB) stars (isotopes 
with A from 90 to 208); the {\it weak }s-component is provided mainly by 
massive stars that explode later on as Supernovae (isotopes with A up to 90). 
The third strong s-component was introduced to reproduce 50\% of $^{208}$Pb
\citep{kap82}. 
The origin of the rapid r-process is still discussed now. The principal 
sources of the r- process nuclei production might be massive stars 
\citep{woo94} or yields from coalescence of two neutron stars (sometimes, 
black hole and neutron star \citealt{fre99}; \citealt{sur08}).
Such s-process elements as Sr, Y, Zr are labeled as {\it light }s-elements (ls);
barium heads the group of {\it heavy }s-elements (hs) (Ba, La, Ce, Nd, Sm) 
(that corresponds to the first and the second peaks of the s-process elemental 
abundance curve).
According to the modern view on the role of the third dredge-up and 
multiplicity of neutron-capture contributions, the AGB stars bring the most 
to the s-elements enrichment at near-solar metallicities (for example, 
\citealt{cri09}).
The estimates of s-process relative contribution to the solar Ba abundance, 
conducted in series of works, are rather similar - from 81\% to 85\% 
(\citealt{kap89}, \citealt{ser09}).
However, the barium abundance variations relative to metallicity draw 
particular attention. As a rule, with diminishing metallicity, a decrease in 
the ratio of Ba abundance relative to iron [Ba/Fe] is observed with an 
 marked dispersion of [Ba/Fe] at [Fe/H] 
within --3 - --4 dex (\citealt{fra07}, \citealt{fre10}). At the indicated 
metallicities, it is an r-process, and therefore massive stars, that contribute 
to the barium enrichment the most (\citealt{arg04}, \citealt{ce06}).
The increasing number of less massive AGB stars with the growing metallicity 
in the Galaxy changes proportions of r- and s-processes contribution to the 
barium enrichment. As a result, at the near-solar metallicity, 
the [Ba/Fe] overabundance relative to the solar ratio is noticed.
The barium abundance in the stars with metallicity from --1 to +0.3 has been 
investigated in a number of studies (\citealt{mag00}, \citealt{mag01},
\citealt{be05}, \citealt{red06}). On considering the dependence of the barium 
abundance on metallicity, the presence of the [Ba/Fe] peak at [Fe/H] of about 
--0.2 dex calls attention; that peak has been discovered earlier in the 
classical research by \citet{edv93}. 
In the recent studies by \citet{be05}, and \citet{red06}, that investigated a 
great number of stars of the thick and thin discs, the lower barium abundance 
in the thick disc stars relative to that in the thin disc members was obtained;
but as per research data by \citet{be05}, the above mentioned peak was noticed 
only in the thin disc stars. As stated by the modern conception, the 
interstellar medium, where dwarf stars with metallicity from --1 to +0.3 dex 
(those are thick and thin discs stars) were born, has been contaminated with the 
products of evolution of several generations of stars; thus, to adequately 
reconstruct the enrichment history, it is necessary to use a Galactic evolution 
model that allows for processes changing the abundance of chemical elements in 
the course of the evolution. It is known that the abundance of barium can 
deviate from the local thermodynamic equilibrium value (\citealt{mas99}, 
\citealt{an09}), especially in metal-poor stars. In the present study, we 
used the NLTE approach to determine more accurately the barium abundance in 
172 dwarf stars that earlier had been kinematically assigned to the thick and 
thin discs \citep{mis04}, and also to specify and analyze its behaviour in the 
stars of the thin and thick Galactic discs in order to compare with the 
Galactic evolution models. 

\section{Observations and selection}

The spectra of 172 stars (F-G-K V) were obtained in the region of  
${\rm \lambda\lambda 4400-6800\,\AA}$ with the signal-to-noise 
ratios  $S/N$ about 100-300 using the 1.93 m telescope at Observatoire de  
Haute-Provence (OHP, France) equipped with the echelle-spectrograph ELODIE 
(\citealt{ba96}). The resolving power R was 42000. 
The initial spectra processing was made online through the telescope 
\citep{kat98}. 
The studied spectra subsequent processing (including the continuous spectrum 
level set up, the development of a dispersion curve, the measurement of 
equivalent widths EW, etc.) was performed using the DECH20 software package 
\citep{gal92}.

To select the stars that belong to the thin and thick discs, we used a 
kinematic approach to assign the target stars by the probability of their 
membership in either the thin or thick discs  on the 
base of their spatial velocities, kinematic parameters of the discs, 
and the percentage of stars in each disc. 
More details can be found in \citet{mis04}, \citet{sob05}.

\section{Determination of parameters}

The atmospheric parameters of the target stars were determined in the previous
studies (\citealt{mik01}, \citealt{mis04}).

The effective temperatures $T_{\rm eff}$ were estimated by the calibration of 
the ratio of the  central depths  of the lines with different potentials of 
the lower bf level in the region of metallicity [Fe/H] $>$ -0.5 dex developed 
by \citet{kov03} with a typical accuracy better than 10 ~K. The effective temperatures 
for the more metal-poor stars ([Fe/H] $<$ -0.5) were determined by the 
fitting of the $H_{\alpha}$ line wings \citep{mik01}
%
%
To testify the temperatures scales of \citet{kov03} and \citet{mik01} we 
have compared in \citet{mis04} of the temperatures determined by us with the 
results of by \citet{AAMR96}, \citet{BLG98}, and \citet{dB98}.
Our temperature scale is slightly hotter than their by $\sim 20-30$ K,
as mentioned in \citet{kov03}, but the dispersion is satisfactory ($\sim 80$ K)

The surface gravities $log(\sl g)$ were determined by two methods for the stars 
with $T_{\rm eff}$ higher than 5000 K (namely, by the iron ionisation balance 
and the parallax). For the cooler stars the parallax method was only used.
\citet{all99} have analysed this two methos and
 concluded that astrometric
and spectroscopic (iron ionization balance) gravities were in good agreement
in the metallicity range $-1.0<$ [FeH] $<+0.3$. We compared \citep{mis04} our 
adopted surface gravities to those determined astrometrically
by \citet{all99} and obtain a mean difference and standard deviation of
-0.01 and 0.15 respectively for 39 stars in common. This is consistent with
an accuracy of 0.1 dex on our estimated spectroscopic gravities.

The microturbulent velocity $V_{\rm t}$ was derived considering that the 
iron abundance log A(Fe) obtained from the given Fe\,{\sc i} line,
is not correlated with the equivalent width EW of that line.

The adopted value of the metallicity [Fe/H] is the iron abundance obtained from
Fe\,{\sc i} lines. As it is known, the neutral iron lines deviate from the 
local thermodynamic equilibrium (LTE) and that consequently affects the iron 
abundances obtained from those lines. However, in the temperature and 
metallicity ranges of the target stars, the NLTE corrections do not exceed 
0.1 dex \citep{mas10}.

The comparison of the determined atmospheric parameters to the data obtained 
by other authors is presented in previous studies \citep{mis04}. The external 
accuracy of the effective temperature is $\Delta$ $T_{\rm eff}$ = $\pm$100 K
and of the surface gravity is $\Delta$ $log(\sl g)$ = $\pm$ 0.2 dex.

\section{Determination of the barium abundance, the model of atom}

The barium abundance in dwarf stars is determined from Ba\,{\sc ii} 4554~\AA,
5853~\AA, 6141~\AA, and 6496~\AA\ lines under the non-LTE approximation.

The used model of the Ba atom is described in detail by \citet{an09}. 
The model contains 31 levels of Ba\,{\sc i}, 101 levels of Ba\,{\sc ii} with
n $<$ 50, and the ground level of Ba\,{\sc iii}. In the detailed 
study, we included 91 bound-bound transitions between the first 28 levels of 
Ba\,{\sc ii} with $n < 12$ and $l < 5$.
%
%
%
To check the completeness of our atomic model we perforemd the test 
calculations. The test calculations that take into account only transitions 
between the first 20 levels of Ba\,{\sc ii} have shown that decrease of the 
number of considered levels practically does not affect the lines of interest.

Some uncertainty of the NLTE analysis of the barium spectrum is caused by the 
lack of information on the photoionization cross sectors for different levels. 
We used the results of the scaled Thomas-Fermi method application \citep{hof79}. 

The effective collision strengths for electron excitation for the transitions 
between the first levels ($\rm 6s^2S$, $\rm 5d^2D$ and $\rm 6p^2P^0$) were used
following to \citet{sb98}. The experimental cross-sections for the 
transitions $\rm6s^2S-7s^2S$ and $\rm6s^2S-6d^2D$ were taken from 
\citet{cra74}. 
The collisional rates for the transitions between sublevels 
$\rm 5d^2D$, $\rm 6p^2P^0$ and $\rm7s^2S$, $\rm 6d^2D$, as well as between 
$\rm7s^2S$ and $\rm6d^2D$ were estimated by the corresponding formula 
from \citet{sob81}. For the rest of the allowed transitions, we applied 
the \citet{reg62} formula, while the Allen's formula \citep{al73} was used for 
the forbidden transitions. The collisional ionization rate from 
the ground level of Ba\,{\sc ii} was computed with the appropriate formula 
from \citet{sob81}.

The collisions with hydrogen atoms were considered using the \citet{sth92} 
formula with a correction factor of 0.1 that was derived empirically by 
analyzing the profiles of the Ba lines in the solar spectrum.
%
%
The change of this coefficient in the range from one to zero does not
affect the resonance line profile in the solar spectrum, while subordinate
lines show variation of the equivalent width in the range 5.5\%-7\%.

\begin{figure}
\resizebox{\hsize}{!}
{\includegraphics{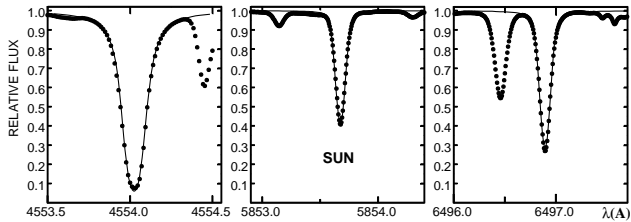}}
\caption[]{Profile fitting in the solar spectrum.}
\label{sun}
\end{figure}

The odd barium isotopes have hyperfine splitting of their levels and thus 
several HFS components for each line \citet{rut78}. 
Therefore, the line 4554 \AA~ and 6496 \AA~ was fitted in the solar spectrum 
(see Fig. \ref{sun}) by adopting the even-to-odd abundance ratio of 
82:18 \citep{cam82}. 
The hyperfine splitting for the lines 5853~\AA~ and 6141~\AA~ is insignificant. 
The parameters of the analyzed lines are given in Table \ref{hfs}. 
We determined $\Gamma_{vdw}$ by fitting the computed line profiles to those 
observed in the solar spectrum. 
%
%
These estimates agree very well with results of theoretical calculations
\citet{bar00} for resonance line (the difference is only 0.05 dex)
and for the lines 6494 and 5141 (the difference is about 0.10 dex).
Nevertheless the 5853 line wings were fitted to observed profile only with
$\Gamma_{vdw} = -7.19$, that is 0.39 dex larger than the value adopted in
\citet{bar00}.  At the same time we took into account that this line is
blended with weak Fe\,{\sc i} line.

The estimates of the radiative broadening for 
the lines 4554, 5853 and 6496~\AA~ are taken from \citet{mab96}; 
for the line 6141~\AA~ - from the Vienna Atomic Line Database (VALD). 
%
%
Calculated barium line profiles in de Solar spectrum \citep{kur84} adgree 
well with observed profiles if we use barium abundanse 2.17 
( in good agreement with the estimation of the solar abundance of barium
of \citet{asp05}).

The NLTE profiles of the barium lines were computed using a modified version of 
the MULTI code \citep{car86}. The modifications applied are described in 
\citet{kor99}. 

The Ba\,{\sc ii} lines are to some extent blended with other metallic lines
(especially the line 6141 \AA), to compare the NLTE barium line profiles to 
those in the observed spectrum, it is necessary to use a combination of the 
NLTE and LTE synthetic spectra. 
It was made with the updated SynthV code \citep{tsy96} that was designed for 
synthetic spectrum calculations under the LTE.
%
%
This code, as well as MULTI code (our version), is based on the opacities
from ATLAS9. The test calculations of the LTE profiles generated by SynthV
and MULTI showed an excellent agreement.

Using the indicated program, we computed synthetic spectra for the selected
regions, comprising the target Ba II lines, on having taken into account all 
lines of each region listed in the VALD. Specifically for the barium lines, 
the corresponding {\it b}-factors (factors of deviation from the LTE level 
populations), computed by the MULTI, were included into the SynthV and 
subsequently applied in calculation of the barium line source function.

For the computations we used the stellar atmosphere models computed using the 
ATLAS9 program with new ODF tables \citep{ck04} without 
overshooting for each star separately.

\begin{figure}
\resizebox{\hsize}{!}
{\includegraphics{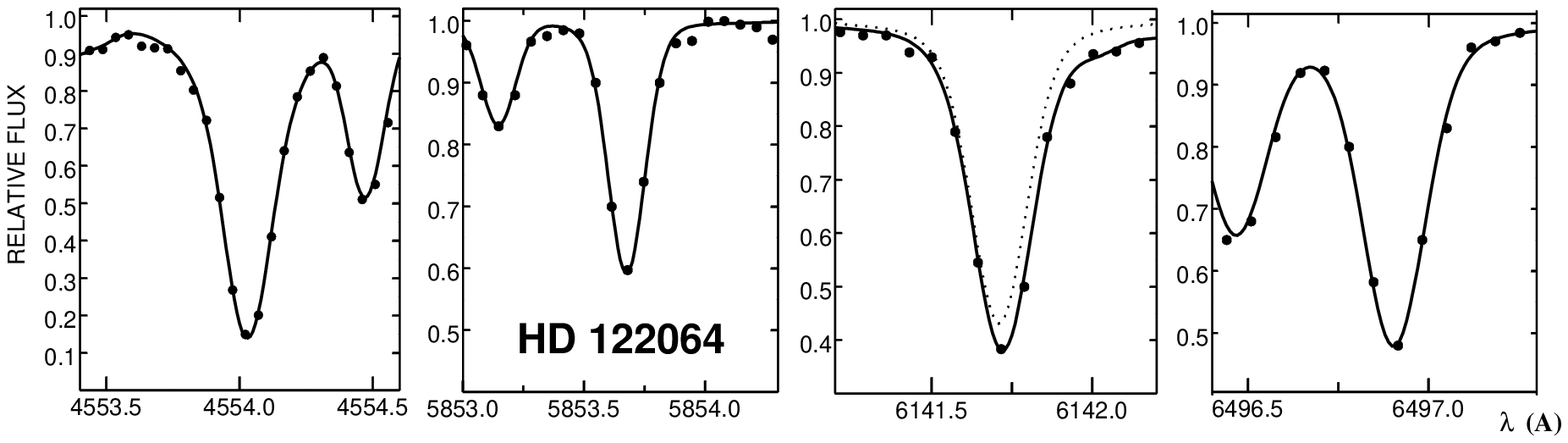}}
\resizebox{\hsize}{!}
{\includegraphics{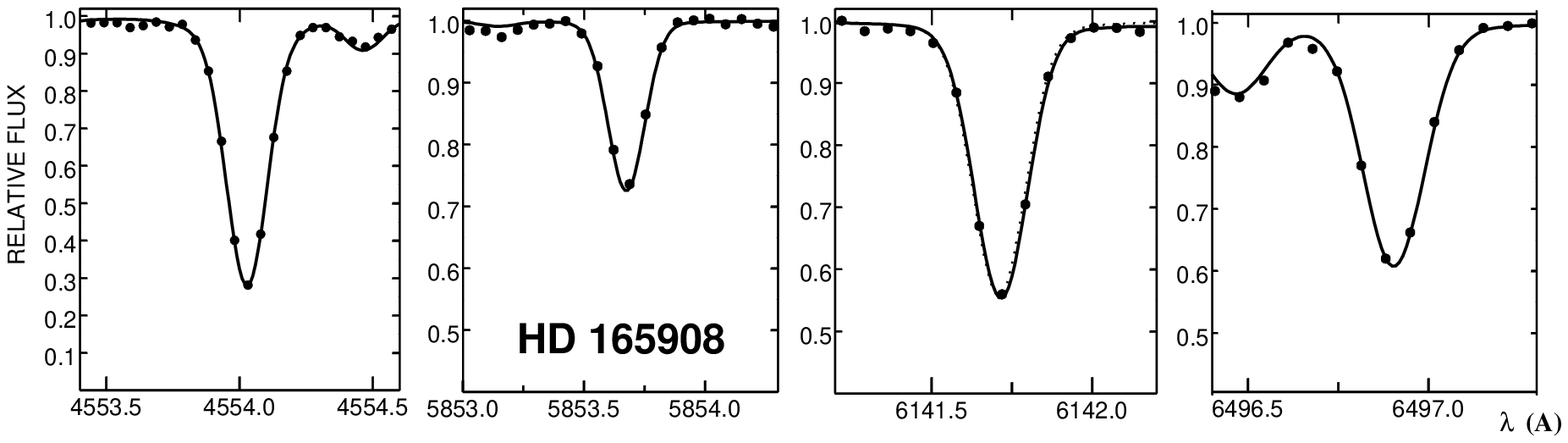}}
\caption[]{Profile fitting in the stellar spectrum. The line 6141 \AA~ is 
blended with the iron line (the dot line - is the barium line profile).}
\label{star}
\end{figure}

The accuracy of determination of the barium abundance in the investigated 
stars, obtained by fitting the synthetic spectra to those observed, is about 
0.03 dex. Some examples of the profile fitting are presented on Fig. \ref{star}. 

\begin {table}
\caption {Parameters of the barium lines.
}
\label {hfs}
\begin{tabular}{l c c c c }
\hline
$\rm \lambda (\AA)$ & HFS & \it f & $\rm log~ \gamma_{rad}$ & $log \Gamma_{vw}$\\
\hline
             &$\Delta\lambda$, m\AA&   solar   &\\
\hline
4554.03&   0&  0.597&  $ 8.20    $&  -7.60\\
       &  18&  0.081& \\
       & -34&  0.049& \\
\\
5853.70&   -&  0.025&  $ 8.20    $&  -7.19\\
\\
6141.71&   -&  0.025&  $ 7.77    $&  -7.47\\
\\
6496.90&   0&  0.086&  $ 8.10    $&  -7.47\\
       &  -4&  0.012&\\
       &   9&  0.007&\\ 
\hline  
\end {tabular}  
\end {table}

The total error of determination of the barium abundance is assumed to be
0.10-0.15 dex with an allowance of errors in estimations of effective 
temperature ($\Delta T_{\rm eff}$ = $\pm$ 100 K), surface gravity 
($\Delta log(\sl g)$ = $\pm$ 0.2), microturbulent velocity 
($\Delta V_{\rm t}$ = $\pm$ 0.2 km/s) and the determined value of 
metallicity ($\Delta$ [FeH] = $\pm$0.10.)

\begin{figure}
\resizebox{\hsize}{!}
{\includegraphics{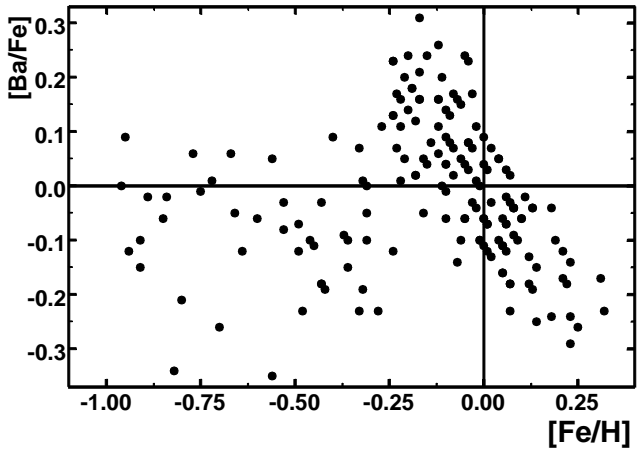}}
\caption[]{The LTE barium abundnace (from the only 4554 \AA~ line)}
\label{lte}
\end{figure}

Firstly, the barium abundance was determined from the only line 4554 \AA~ under 
the LTE. The obtained results are given in Fig. \ref{lte}. The shape of the 
obtained dependence between [Ba/Fe] and [Fe/H] appeared rather strange. A 
sharp peak and an intrinsic dispersion of the barium abundance ratio can be 
noticed.

\begin{figure}
\resizebox{\hsize}{!}
{\includegraphics{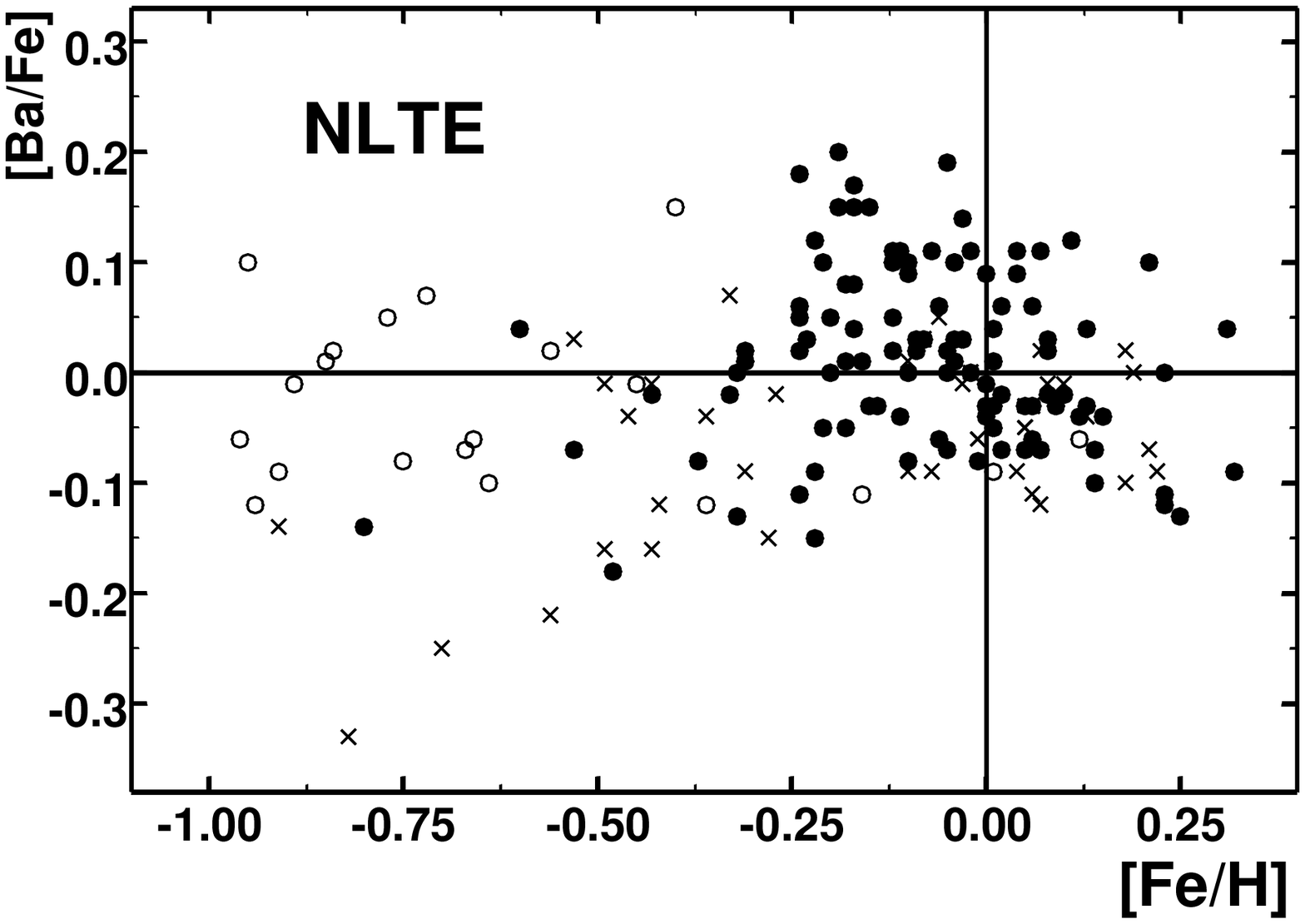}}
\resizebox{\hsize}{!}
{\includegraphics{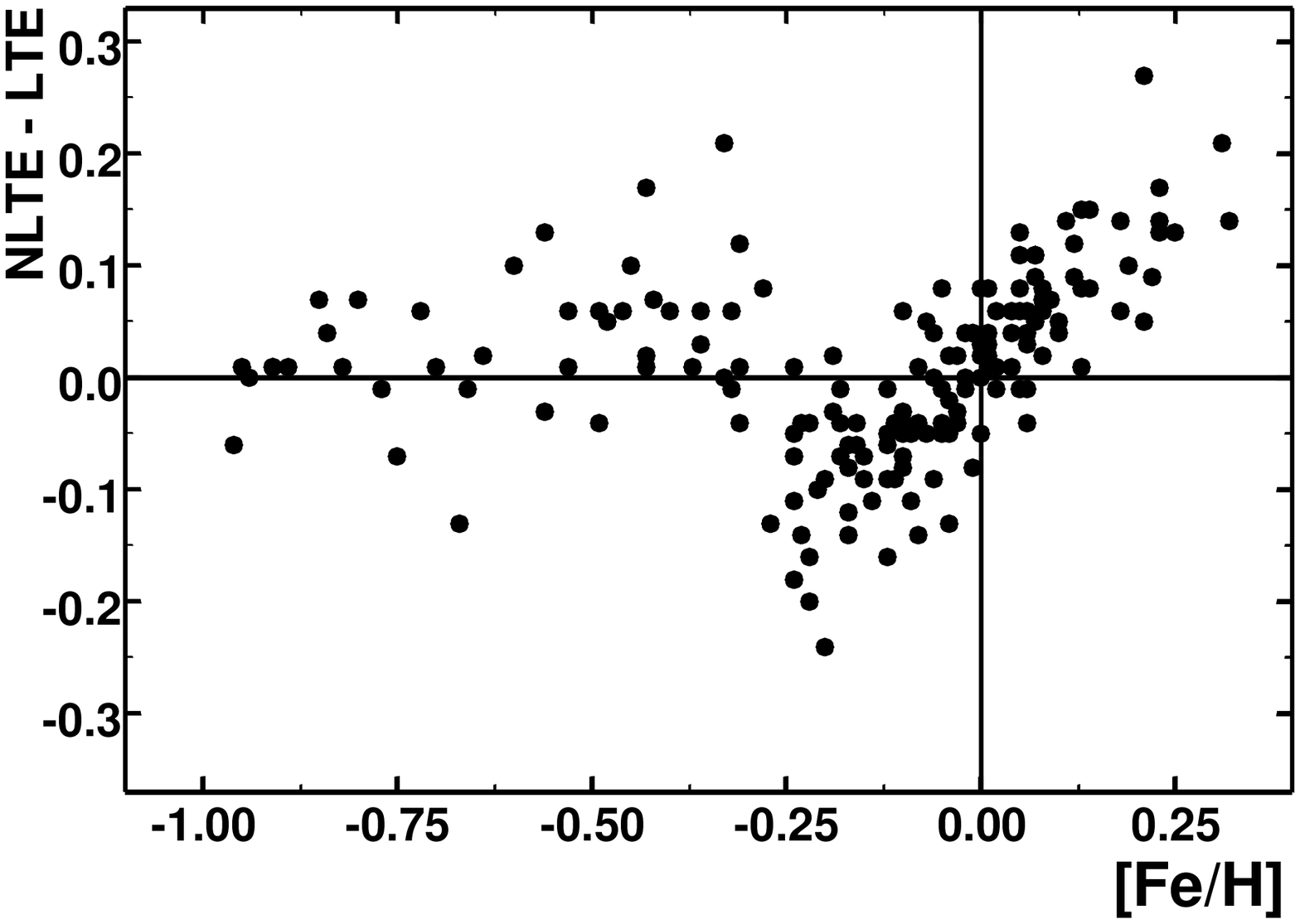}}
\caption[]{The NLTE barium abundnace and the NLTE corrections vs. metallicity.
The thin disc stars are marked as filled circles, the thick disc stars - 
as open circles,  the others - as crosses.}
\label{nlte}
\end{figure}

The dispersion of the obtained barium abundance ratios was appreciably 
decreased by considering deviations from the LTE and using the four 
Ba\,{\sc ii} lines; however, a small [Ba/Fe] overabundance at [Fe/H] = --0.2 
dex still remained. The results are given in Table. \ref{tabnlte} and  
Fig. \ref{nlte}. 
The difference between the LTE and NLTE abundance determination is shown in 
the same figure. 
%
%
These differences equally conditioned by the two circumstances, firstly,
by the use of four lines instead of one, and secondly, by the differences
in the LTE and NLTE determinations. Let us note that the way of presentation
of the barium abundance results in Fig. \ref{lte} (inclined parallel lines) is 
caused by the step discontinuity in synthetic spectrum fitting.

In Fig. \ref{nlte}, the thin disc stars are marked as filled 
circles, the thick disc stars - as open circles,  the others - as crosses. 
Thus, it is evident that the LTE 
analysis of the only one 4554~\AA~ line leads to the occurrence of systematic 
errors.

The results of the comparison of the obtained barium abundances and the 
atmospheric parameters to those of the other authors for common stars are 
given in Table \ref{comp}, where $\Delta$ $T_{\rm eff}$ - the average effective
temperature difference, $\Delta$ $log(\sl g)$ - the average surface gravity 
difference, $\Delta$ [Fe/H] - the average metallicity difference, n - the 
number of common stars.

\begin {table}
\caption {The results of the comparison of the obtained Ba  NLTE abundance 
to the 
data from the LTE determination by  \citet{be05} - B05, \citet{red06} - 
R06, and NLTE determination \citet{maz06} - M06 for common stars.}
\label {comp}
\begin{tabular}{l c c c c c}
\hline
            & $\Delta$ $T_{\rm eff}$ &$\Delta$ $log(\sl g)$ & $\Delta$ [Fe/H] & $\Delta$ [Ba/Fe]   &  n \\
\hline
B05    &     19 $\pm$76  &  --0.09$\pm$0.19  &  --0.03$\pm$0.08  &  --0.06$\pm$0.07   &  10\\
            &          &          &          &           &    \\
R06     &     92 $\pm$29  &  --0.20$\pm$0.24  &  --0.01$\pm$0.04  &    0.12$\pm$0.11   &   8\\
            &          &          &          &           &    \\
M06      &     12 $\pm$64  &  --0.11$\pm$0.22  &    0.01$\pm$0.07  &    0.06$\pm$0.08   &  13\\
\hline  
\end {tabular}  
\end {table}   

As is obvious from the mentioned comparison, the obtained values corroborate 
the external accuracy for $T_{\rm eff}$ and $log(\sl g)$ declared by 
\citet{mis04}; and they are rather harmonized for metallicity (within 0.03 dex). 
The barium abundance ratio difference $\Delta$ [Ba/Fe] = 0.12, obtained by 
comparing to the ratio by \citet{red06} is greater than that received by 
collating with the other studies; it could be caused by dissimilarity of 
temperatures and gravities, and maybe by some disparateness of the NLTE and 
LTE approaches applied in those studies. At the same time, such a shift is 
within the accuracy limits for the barium abundance determination. 

\section{Discussion of the results obtained}

The obtained barium abundance values are more accurate in view of the NLTE 
approach application and derivation of stellar models considering distinct 
chemical compositions of each star; those values are used to analyze the barium
abundance in the stars of the thin and thick Galactic discs, assigned by 
kinematic criteria. The average barium abundances for the thin and thick discs 
are 0.01 $\pm$0.08 and --0.03 $\pm$0.07, respectively. The average difference 
is 0.04 dex. 
To estimate the significance of the difference obtained, we consider the barium 
abundance variations relative to magnesium - an element that explicitly shows 
the dissimilarity of abundances in the thick and thin discs 
(\citealt{gra00}, \citealt{fur98}, \citealt{mag01}, \citealt{be05},
\citealt{red06}, \citealt{mis04}, etc.) and has the only source of production, 
namely Supernovae of 8-12 $M_{\sun}$. To make such a comparison, we can use 
the magnesium abundances determined under the NLTE approximation in the 
previous studies \citep{mis04} that specified the assignment to the discs.

\begin{figure}
\resizebox{\hsize}{!}
{\includegraphics{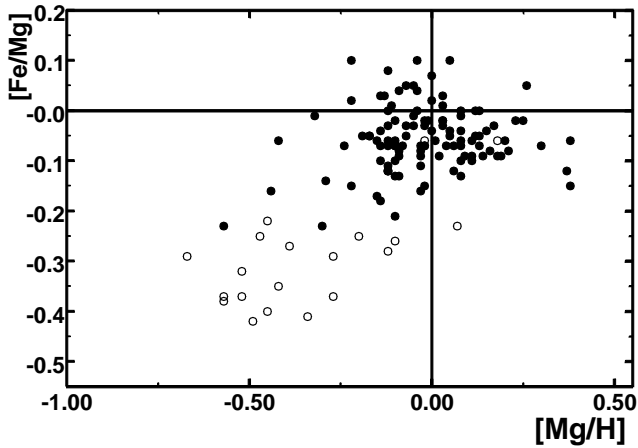}}
\caption[]{[Fe/Mg] vs. [Fe/H]. The stars of the thin and thick disks are 
marked as filled and open circles, respectively.}
\label{femg}
\end{figure}

\begin{figure}
\resizebox{\hsize}{!}
{\includegraphics{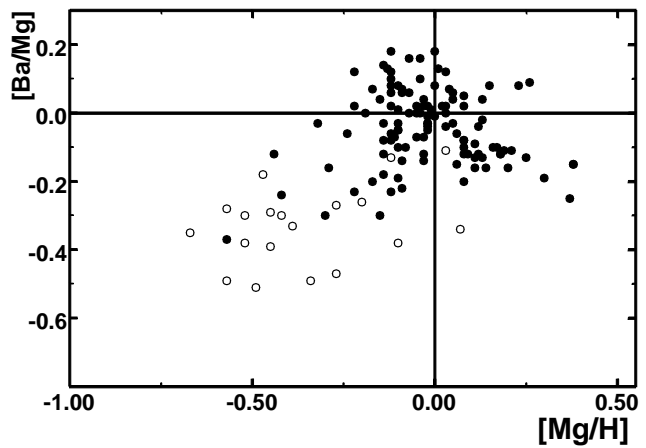}}
\caption[]{[Ba/Mg] vs. [Mg/H]. The stars of the thin and thick disks are 
marked as filled and open circles, respectively.}
\label{bamg}
\end{figure}

The relative abundances [Fe/Mg] vs. [Mg/H] and [Ba/Mg] vs. [Mg/H] only for the 
stars of the thick and thin discs are presented in Fig. \ref{femg}, \ref{bamg}.
The dissimilarity between the magnesium abundances in the thick and thin discs 
is clearly retraced in Fig. \ref{femg} that shows the correlation between 
[Fe/Mg] and [Mg/H]. For the dependence of [Ba/Mg] relative to [Mg/H] shown in 
Fig. \ref{bamg}, we observe a barely noticeable trend with the magnesium 
abundance increasing; the barium underabundance in the thick disc relative to 
the solar ratio and a "cloud" in the thin disc. That means that the difference 
in the barium behaviour in the thick and thin discs is retraced more clearly. 
However, the above may also be evidence of different sources of production of 
those elements. It should be noted that such comparison is rather relative in 
this case as the elemental enrichment in the disc stars is contributed by the 
reproduction by several generations of stars; i.e. there is enough time for the 
interstellar medium to get enriched with elements produced by numerous 
predecessors either massive or less massive, including Supernovae and the AGB 
stars, whose fractions to the enrichment are not proportionate.

Therefore, to adequately reconstruct the enrichment history, it is necessary 
to use chemical evolution models that allow for many factors provoking changes 
in the abundance of one or another chemical element, e.g. elemental production 
sources, the Galactic structure model, the initial mass function, the star 
formation rate, etc.

In this work, we are to compare the obtained barium abundances to the 
computations of the model by \citet{ser09}. 

It is interesting to make such a comparison by considering several points:  
1) what is the s- and r-processes contribution to the barium enrichment at 
metallicities from -1 to +0.3; 
2) what stars (sources) contribute to the barium production the most; 
3) whether the barium enrichment in the thick disc differs from that in the 
thin disc within the bounds of the applied model of the thick disc formation 
and the Galaxy formation as a whole. In the study by \citet{ser09}, firstly, 
using the FRANEC code \citep{chi98}, the s-process contribution to each 
investigated isotope at the solar system formation epoch was computed following 
the AGB stars yield only. Then, using the r-process residual method, the 
r-process fraction was determined for each isotope by subtracting the 
corresponding s-process fraction; subsequently, the chemical evolution model 
was recomputed. For example, considering barium, the r-residual of 21\% was 
assumed \citep{tra99}, the main contributors were Supernovae of type II with 
masses within 8 $<$ $M/M_{\sun}$ $<$ 12. The calculations were made on the base of 
the model by \citep{fer92}, that allows of the Galaxy division into three 
components: the halo, the thin and thick discs. In so doing, it was assumed 
that the thin disc is formed from the matter falling from the halo and the 
thick disc. The problem of the above model is the fact that the thick disc can 
not be formed from the halo gas \citep{wg92}. It is not possible to distinguish
the thin and thick discs in the two-infall model by \citet{chi97} either. 
Therefore, taking into account computations of the Galactic evolution models as
of today, we can not clarify the question on the difference of the Galactic 
discs enrichment with one or another element.

\begin{figure}
\resizebox{\hsize}{!}
{\includegraphics{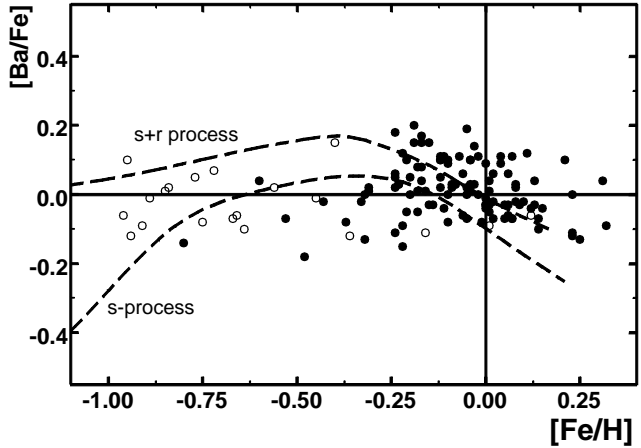}}
\caption[]{The results of the comparison of the observed trend of [Ba/Fe] vs. 
[Fe/H] to the computations of the Galactic chemical evolution (dashed lines) 
by \citet{ser09}.}
\label{rs}
\end{figure}

Using the model by \citet{ser09}, we can determine the s- and r-processes 
contributions to the barium enrichment with allowance of the s-elements (in 
particular, barium) production sources applied in the model. It is indicated 
that the s-elements enrichment at the solar metallicity is contributed mainly 
by the AGB low-mass stars (1.5 $<$ $M/M_{\sun}$ $<$ 3). The results of the 
comparison of the observed correlation between the barium abundance and 
metallicity to the calculations of the model by \citet{ser09} are shown in 
Fig. \ref{rs}.

It is evident that the observations are poorly specified by the estimated 
r+s process curve for the thin disc stars with [Fe/H]$>$0.1 dex and 
[Fe/H] $<$ --0.2 dex. That means that the r-process contribution differs from 
that fraction applied in the model by \citet{ser09}.

The given comparison indicates that the barium enrichment in the thick and thin
discs (at metallicity from -1 to +0.3) is a process more complicated than that 
specified in the considered model; and it requires further studying.

\section{Conclusions}

1. The barium abundance in 172 dwarf stars in the metallicity range from 
--1 to +0.3 is determined from four lines under the NLTE approximation with 
distinct atmospheric models.

2. A slight difference (0.04 dex) in the average barium abundances is indicated
between the investigated stars, kinematically assigned to the thick (21 stars) 
and thin (109 stars) discs. However, the dependence of [Ba/Mg] vs. [Mg/H] has 
shown different behaviour of the barium abundance in the thick and thin discs. 

3. The comparison to the Galactic chemical evolution model by \citet{ser09}
has indicated that the barium enrichment is evidently specified by 
contributions of the s- and r-processes, applied in the model for the stars 
with metallicity --0.2 $<$[Fe/H] $<$0.1 dex. However, the r-process contribution in
the stars with other metallicity ranges differs from that fraction applied in 
the Serminato model.

\begin{table}
\caption[]{ Atmospheric parameters and Ba$_{\rm NLTE}$ abundances}
\label{tabnlte}
\begin{tabular}{ccccc}
\hline
\hline
HD& $T_{\rm eff}$ & $log(\sl g)$ & [FeH] & [Ba/H]$_{\rm NLTE}$ \\
\hline
 HD000245   &5400&3.4&--0.84& --0.82 \\
 HD001562   &5828&4.0&--0.32& --0.32 \\
 HD001835   &5790&4.5& +0.13& +0.17  \\
 HD003765   &5079&4.3& +0.01& --0.08 \\
 HD004307   &5889&4.0&--0.18& --0.10 \\
 HD004614   &5965&4.4&--0.24& --0.22 \\
 HD005294   &5779&4.1&--0.17& --0.02 \\
 HD006582   &5240&4.3&--0.94& --1.06 \\
 HD008648   &5790&4.2& +0.12& +0.08  \\
 HD009826   &6074&4.0& +0.10& +0.08  \\
 HD010145   &5673&4.4&--0.01& --0.07 \\
 HD010307   &5881&4.3& +0.02& +0.00  \\
 HD010476   &5242&4.3&--0.05& --0.05 \\
 HD010780   &5407&4.3& +0.04& +0.13  \\
 HD011007   &5980&4.0&--0.20& --0.15 \\
 HD013403   &5724&4.0&--0.31& --0.40 \\
 HD013507   &5714&4.5&--0.02& +0.09  \\
 HD013783   &5350&4.1&--0.75& --0.83 \\
 HD013974   &5590&3.8&--0.49& --0.50 \\
 HD014374   &5449&4.3&--0.09& --0.07 \\
 HD017674   &5909&4.0&--0.14& --0.17 \\
 HD017925   &5225&4.3&--0.04& --0.01 \\
 HD019019   &6063&4.0&--0.17& +0.00  \\
 HD019308   &5844&4.3& +0.08& +0.07  \\
 HD019373   &5963&4.2& +0.06& +0.03  \\
 HD022049   &5084&4.4&--0.15& +0.00  \\
 HD022484   &6037&4.1&--0.03& +0.00  \\
 HD022556   &6155&4.2&--0.17& --0.13 \\
 HD022879   &5972&4.5&--0.77& --0.72 \\
 HD023050   &5929&4.4&--0.36& --0.40 \\
 HD024053   &5723&4.4& +0.04& +0.15  \\
 HD024206   &5633&4.5&--0.08& --0.05 \\
 HD028005   &5980&4.2& +0.23& +0.23  \\
 HD028447   &5639&4.0&--0.09& --0.06 \\
 HD029150   &5733&4.3& +0.00& --0.03 \\
 HD029310   &5852&4.2& +0.08& +0.10  \\
 HD029645   &6009&4.0& +0.14& +0.07  \\
 HD030495   &5820&4.4&--0.05& +0.14  \\
 HD030562   &5859&4.0& +0.18& +0.20  \\
 HD032147   &4945&4.4& +0.13& +0.09  \\
 HD033632   &6072&4.3&--0.24& --0.06 \\
 HD034411   &5890&4.2& +0.10& +0.09  \\
 HD038858   &5776&4.3&--0.23& --0.20 \\
 HD039587   &5955&4.3&--0.03& +0.11  \\
 HD040616   &5881&4.0&--0.22& --0.10 \\
 HD041330   &5904&4.1&--0.18& --0.17 \\
 HD041593   &5312&4.3&--0.04& +0.06  \\
 HD043587   &5927&4.1&--0.11& --0.15 \\
 HD043856   &6143&4.1&--0.19& --0.04 \\
 HD043947   &6001&4.3&--0.24& --0.18 \\
 HD045067   &6058&4.0&--0.02& --0.02 \\
 HD050281   &4712&3.9&--0.20& --0.20 \\
 HD051419   &5746&4.1&--0.37& --0.45 \\
 HD055575   &5949&4.3&--0.31& --0.29 \\
 HD058595   &5707&4.3&--0.31& --0.30 \\
 HD061606   &4956&4.4&--0.12& --0.10 \\
 HD062613   &5541&4.4&--0.10& --0.10 \\
 HD064606   &5250&4.2&--0.91& --1.05 \\
 HD064815   &5864&4.0&--0.33& --0.26 \\
 HD065583   &5373&4.6&--0.67& --0.74 \\
 HD065874   &5936&4.0& +0.05& --0.02 \\
 HD066573   &5821&4.6&--0.53& --0.60 \\
 HD068017   &5651&4.2&--0.42& --0.54 \\
 HD068638   &5430&4.4&--0.24& --0.19 \\
\hline                         
\end{tabular}                  
\end{table}                    

\begin{table}
\contcaption{}
\begin{tabular}{ccccc}
\hline
\hline
HD& $T_{\rm eff}$ & $log(\sl g)$ & [FeH] & [Ba/H]$_{\rm NLTE}$ \\
\hline
 HD070923   &5986&4.2& +0.06& +0.00  \\
 HD071148   &5850&4.2& +0.00& --0.01 \\
 HD072760   &5349&4.1& +0.01& +0.05  \\
 HD072905   &5884&4.4&--0.07& +0.04  \\
 HD073344   &6060&4.1& +0.08& +0.06  \\
 HD075732   &5373&4.3& +0.25& +0.12  \\
 HD076151   &5776&4.4& +0.05& +0.02  \\
 HD076932   &5840&4.0&--0.95& --0.85 \\
 HD081809   &5782&4.0&--0.28& --0.43 \\
 HD082106   &4827&4.1&--0.11& +0.00  \\
 HD088072   &5778&4.3& +0.00& --0.03 \\
 HD089251   &5886&4.0&--0.12& --0.07 \\
 HD089269   &5674&4.4&--0.23& --0.20 \\
 HD091347   &5931&4.4&--0.43& --0.45 \\
 HD095128   &5887&4.3& +0.01& --0.04 \\
 HD098630   &6060&4.0& +0.22& +0.13  \\
 HD101177   &5932&4.1&--0.16& --0.15 \\
 HD102870   &6055&4.0& +0.13& +0.10  \\
 HD106516   &6165&4.4&--0.72& --0.65 \\
 HD107213   &6156&4.1& +0.07& +0.09  \\
 HD107705   &6040&4.2& +0.06& +0.12  \\
 HD108954   &6037&4.4&--0.12& --0.01 \\
 HD109358   &5897&4.2&--0.18& --0.23 \\
 HD110833   &5075&4.3& +0.00& --0.04 \\
 HD110897   &5925&4.2&--0.45& --0.46 \\
 HD112758   &5203&4.2&--0.56& --0.78 \\
 HD114710   &5954&4.3& +0.07& +0.18  \\
 HD115383   &6012&4.3& +0.11& +0.23  \\
 HD116443   &4976&3.9&--0.48& --0.66 \\
 HD117043   &5610&4.5& +0.21& +0.31  \\
 HD117176   &5611&4.0&--0.03& --0.04 \\
 HD117635   &5230&4.3&--0.46& --0.50 \\
 HD119802   &4763&4.0&--0.05& --0.03 \\
 HD122064   &4937&4.5& +0.07& +0.00  \\
 HD125184   &5695&4.3& +0.31& +0.35  \\
 HD126053   &5728&4.2&--0.32& --0.45 \\
 HD131977   &4683&3.7&--0.24& --0.35 \\
 HD135204   &5413&4.0&--0.16& --0.27 \\
 HD135599   &5257&4.3&--0.12& --0.02 \\
 HD137107   &6037&4.3& +0.00& +0.09  \\
 HD139323   &5204&4.6& +0.19& +0.19  \\
 HD139341   &5242&4.6& +0.21& +0.14  \\
 HD140538   &5675&4.5& +0.02& +0.08  \\
 HD141004   &5884&4.1&--0.02& --0.02 \\
 HD144287   &5414&4.5&--0.15& --0.18 \\
 HD144579   &5294&4.1&--0.70& --0.95 \\
 HD145675   &5406&4.5& +0.32& +0.23  \\
 HD146233   &5799&4.4& +0.01& +0.02  \\
 HD149661   &5294&4.5&--0.04& --0.03 \\
 HD151541   &5368&4.2&--0.22& --0.37 \\
 HD152391   &5495&4.3&--0.08& --0.05 \\
 HD154345   &5503&4.3&--0.21& --0.26 \\
 HD154931   &5910&4.0&--0.10& --0.09 \\
 HD157089   &5785&4.0&--0.56& --0.54 \\
 HD158633   &5290&4.2&--0.49& --0.65 \\
 HD159062   &5414&4.3&--0.40& --0.25 \\
 HD159222   &5834&4.3& +0.06& +0.03  \\
 HD159482   &5620&4.1&--0.89& --0.90 \\
 HD159909   &5749&4.1& +0.06& --0.05 \\
 HD160346   &4983&4.3&--0.10& --0.18 \\
 HD161098   &5617&4.3&--0.27& --0.29 \\
 HD164922   &5392&4.3& +0.04& --0.05 \\
 HD165173   &5505&4.3&--0.05& --0.12 \\
\hline                         
\end{tabular}                  
\end{table}                    

\begin{table}
\contcaption{}
\begin{tabular}{ccccc}
\hline
\hline
HD& $T_{\rm eff}$ & $log(\sl g)$ & [FeH] & [Ba/H]$_{\rm NLTE}$ \\
\hline
 HD165341   &5314&4.3&--0.08& --0.05 \\
 HD165401   &5877&4.3&--0.36& --0.48 \\
 HD165476   &5845&4.1&--0.06& --0.12 \\
 HD165670   &6178&4.0&--0.10& +0.00  \\
 HD165908   &5925&4.1&--0.60& --0.56 \\
 HD166620   &5035&4.0&--0.22& --0.31 \\
 HD168009   &5826&4.1&--0.01& --0.07 \\
 HD173701   &5423&4.4& +0.18& +0.08  \\
 HD176841   &5841&4.3& +0.23& +0.11  \\
 HD182488   &5435&4.4& +0.07& +0.00  \\
 HD182736   &5430&3.7&--0.06& --0.01 \\
 HD183341   &5911&4.3&--0.01& --0.09 \\
 HD184499   &5750&4.0&--0.64& --0.74 \\
 HD184768   &5713&4.2&--0.07& --0.16 \\
 HD185144   &5271&4.2&--0.33& --0.35 \\
 HD186104   &5753&4.2& +0.05& +0.00  \\
 HD186408   &5803&4.2& +0.09& +0.06  \\
 HD186427   &5752&4.2& +0.02& --0.05 \\
 HD187897   &5887&4.3& +0.08& +0.11  \\
 HD189087   &5341&4.4&--0.12& --0.02 \\
 HD190360   &5606&4.4& +0.12& +0.06  \\
 HD190404   &4963&3.9&--0.82& --1.15 \\
 HD191533   &6167&3.8&--0.10& --0.01 \\
 HD194598   &5890&4.0&--1.21& --1.25 \\
 HD195005   &6075&4.2&--0.06& +0.00  \\
 HD195104   &6103&4.3&--0.19& +0.01  \\
 HD197076   &5821&4.3&--0.17& --0.09 \\
 HD199960   &5878&4.2& +0.23& +0.12  \\
 HD201889   &5600&4.1&--0.85& --0.84 \\
 HD201891   &5850&4.4&--0.96& --1.02 \\
 HD202108   &5712&4.2&--0.21& --0.11 \\
 HD203235   &6071&4.1& +0.05& +0.00  \\
 HD204521   &5809&4.6&--0.66& --0.72 \\
 HD205702   &6020&4.2& +0.01& --0.02 \\
 HD208906   &5965&4.2&--0.80& --0.94 \\
 HD210667   &5461&4.5& +0.15& +0.11  \\
 HD210752   &6014&4.6&--0.53& --0.50 \\
 HD211472   &5319&4.4&--0.04& +0.06  \\
 HD215065   &5726&4.0&--0.43& --0.59 \\
 HD215704   &5418&4.2& +0.07& --0.05 \\
 HD217014   &5778&4.2& +0.14& +0.04  \\
 HD218209   &5705&4.5&--0.43& --0.44 \\
 HD219134   &4900&4.2& +0.05& +0.02  \\
 HD219396   &5733&4.0&--0.10& --0.19 \\
 HD224930   &5300&4.1&--0.91& --1.00 \\
\hline                         
\end{tabular}                  
\end{table}

\section*{Acknowledgments}

This work was supported by the Swiss National Science Foundation, project
SCOPES No. IZ73Z0-128180.
We thank the referee M. Asplund for valuable comments that improved the paper.

\label{lastpage}

\bsp

\end{document}